\documentclass{amsart}
\usepackage{fullpage, url, amsmath, amssymb, graphicx}
\usepackage{lscape, afterpage, threeparttable}
\title{Bayesian Prediction for \emph{The Winds of Winter}}
\author{Richard Vale}
\address{University of Canterbury, Private Bag 4800, Christchurch, New Zealand}
\date{31 August 2014}
\begin{document}
\begin{abstract}
Predictions are made for the number of chapters told from the point of view of each character in the next two novels in George R. R. Martin's \emph{A Song of Ice and Fire} series by fitting a random effects model to a matrix of point-of-view chapters in the earlier novels using Bayesian methods.
{\textbf{SPOILER WARNING: readers who have not read all five existing novels in the series should not read further, as major plot points will be spoiled, starting with Table \ref{M}.}}
\end{abstract}
\maketitle
\section{Description of problem}
\subsection{}Five books have so far been published in George R. R. Martin's popular \emph{Song of Ice and Fire} series \cite{AGOT}, \cite{ACOK}, \cite{ASOS}, \cite{AFFC}, \cite{ADWD}. It is widely anticipated that two more will be published, the first of which \cite[foreword]{AFFC} will be entitled \emph{The Winds of Winter} \cite{TWOW}. Each chapter of the existing books is told from the point of view of one of the characters. So far, ignoring prologue and epilogue chapters in some of the books, $24$ characters have had chapters told from their point of view. A chapter told from the point of view of a particular character $x$ will be called a \emph{POV chapter for} $x$. A character who has at least one POV chapter in the series will be called a \emph{POV character.}

\subsection{}The goal is to predict how many POV chapters each of the existing POV characters will have in the remaining two books (especially \cite{TWOW}.) This varies from character to character, since some major characters have been killed off and are unlikely to appear in future novels, whereas other characters are of minor importance and may or may not have chapters told from their point of view.

\subsection{}No attempt is made to deal with characters who have not yet appeared as POV characters. This issue is discussed further in Section \ref{newcharacters}.

\subsection{The Data. }The data consist of a $24 \times 5$ matrix $M$ obtained from \url{http://www.lagardedenuit.com/}, a French fansite. The rows of $M$ correspond to POV characters and the columns to the existing books in order of publication. The $(i,j)$--entry of $M$ is the number $M_{ij}$ of POV chapters for character $i$ in book $j$. The data are displayed in Table \ref{M}.

\begin{table}[!ht]
\centering
\begin{tabular}{lccccc}
character & AGOT & ACOK & ASOS & AFFC & ADWD\\
\hline
Eddard & 15 & 0 & 0 & 0 & 0\\
Catelyn & 11  &  7  &  7 &   0 &   0\\
Sansa & 6&8 &7 &3 &0 \\
Arya &5 &10 &13 &3 &2 \\
Bran &7 &7 &4 &0 &3 \\
Jon Snow &9 &8 &12 &0 &13 \\
Daenerys &10 &5 &6 &0 &10 \\
Tyrion &9 &15 &11 &0 &12 \\
Theon &0 &6 &0 &0 &7\\
Davos &0 &3 &6 &0 &4 \\
Samwell &0 &0 &5 &5 &0 \\
Jaime &0 &0 &9 &7 &1 \\
Cersei &0 &0 &0 &10 &2 \\
Brienne &0 &0 &0 &8 &0 \\
Areo &0 &0 &0 &1 &1 \\
Arys &0 &0 &0 &1 &0 \\
Arianne &0 &0 &0 &2 &0 \\
Asha &0 &0 &0 &1 &3 \\
Aeron &0 &0 &0 &2 &0 \\
Victarion &0 &0 &0 &2 &2 \\
Quentyn &0 &0 &0 &0 &4 \\
Jon Connington &0 &0 &0 &0 &2 \\
Melisandre &0 &0 &0 &0 &1 \\
Barristan &0 &0 &0 &0 &4
\end{tabular}
\vskip 2ex
\caption{The data. Novel titles are abbreviated using their initials, so for example AGOT = \emph{A Game of Thrones} etc. Obtained from \protect\url{http://www.lagardedenuit.com/wiki/index.php?title=Personnages_PoV}. A similar table appears at \protect\url{http://en.wikipedia.org/wiki/A_Song_of_Ice_and_Fire}.}
\label{M}
\end{table}

\subsection{Point and interval predictions.} It is easy to predict how many POV chapters certain characters will have in the next book. For example, if character $x$ was beheaded in book $1$ and has not appeared in subsequent books, most readers would predict that $x$ will have $0$ POV chapters in book $6$. If we denote by $X_{x6}$ the number of POV chapters for $x$ in book $6$, we predict $X_{x6}=0$. The prediction $X_{x6}=0$ is a \emph{point prediction.} On the other hand, suppose it is believed that a certain character $y$ will have $9$ POV chapters in book $6$. It is quite plausible that $y$ might have $8$ or $10$ POV chapters instead. On the other hand, it may be thought unlikely that $y$ will appear in $0$ (too few) or $30$ (too many) POV chapters. Instead of the point prediction $X_{y6}=9$, it would be better to give a range of likely values for $X_{y6}.$ This is an \emph{interval prediction.} For example, to say that the interval $[7,9]$ is an $80\%$ credible interval for $X_{y6}$ is to say that there is an $80\%$ probability that $7 \le X_{y6} \le 9$.

\subsection{Probabilistic prediction. }\label{probabilisticprediction} Even more refined than an interval of likely values is a probability distribution over all possible values of the number of POV chapters. For example, using $P()$ to denote probability, we could say $P(X_{y6}=9)=0.5$, $P(X_{y6}=10)=P(X_{y6}=8)=0.2$, $P(X_{y6}=7)=0.1$, and $P(X_{y6}=\text{ other values})=0$. This probability distribution completely describes our belief about $X_{y6}$. Conceptually, it can be thought of as a guide for betting. For example, if our beliefs are correct then it is an even money bet that $X_{y6}=9$ and also an even money bet that $X_{y6} \in \{7, 8, 10\}$. Our aim is to give such a probability distribution for each POV character $y$.

\subsection{Modelling. } How can we get a probabilistic prediction as described in Section \ref{probabilisticprediction}? One way would be to assign probabilities based on our gut feelings about the probability of different outcomes, as a traditional bookmaker might do. (An uncharitable reader might even argue that this would be better than the approach taken below.) An alternative approach is to choose a statistical \emph{model}, that is, a process which could plausibly have generated the observed data. Once the model has been chosen, it can be used to predict future data. In Section \ref{model}, we describe a family of possible models which depend on six parameters. Values of these parameters which are likely to have produced the observed data are found, and these parameters are used to generate predictions for future books. This step is repeated many times to build up probability distributions for the predictions. The whole process of finding values of the parameters is called \emph{inference} or \emph{fitting the model.}

\subsection{} In general, the best predictions are obtained by a combination of modelling and common sense. Here we focus entirely on the modelling side and leave common sense behind. The question to be answered could be expressed as: ``What could be predicted about future books if we knew nothing about the existing books except for Table \ref{M}?"

\section{The Model}
\subsection{Description of model. }\label{descriptionofmodel} Denote the number of POV chapters for character $i$ in book $t$ by $X_{it}$, $t \in \{1, 2, 3, 4, 5, 6, 7\}$. We assume that there are times $t_0$ and $t_1$ such that the character is `on-stage' between $t_0$ and $t_1$ and `off-stage' at other times. For example, the character might be killed off at time $t_1$. In other words, $X_{it}=0$ for $t<t_0$ and $t>t_1$. For $t_0 \le t \le t_1$ we assume that POV chapters follow a Poisson distribution with parameter $\lambda_i$.

\subsection{}It is inconvenient to have $t_0$ and $t_1$ as model parameters, so instead we assume that there are $\tau_i \ge 0$ and $\beta_i$ such that
$$
X_{it} \sim
\begin{cases}
\mathrm{Pois}(\lambda_i) &\qquad\text{if } |t-\beta_i| < \tau_i\\
0 &\qquad \text{otherwise.}
\end{cases}
$$
This is the same as putting $t_0 = \beta_i -\tau_i$ and $t_1=\beta_i + \tau_i$ in Section \ref{descriptionofmodel}. Note that we allow $\tau_i$ and $\beta_i$ to take real values, even though $t$ is constrained to be an integer.

\subsection{}It is undesirable for the $\tau_i$, $\beta_i$ and $\lambda_i$ for each character to be independent of the other characters as this would give a model with $3N$ parameters where $N$ is the number of characters. It is unlikely that good predictions could be made from a model with too many parameters. To cut down the number of parameters, we assume that $\lambda_i$, $\beta_i$ and $\tau_i$ are \emph{random effects,} which means that they are samples from some underlying probability distribution. One motivation behind this assumption is that the parameters for different characters are assumed to have something in common. For example, there might be a typical value for $\tau_i$ which reflects how long the average character is likely to last in \emph{A Song of Ice and Fire.}

\subsection{}The $\log(\lambda_i)$ are assumed to be normally distributed and $\beta_i$ and $\tau_i$ are also assumed to be normally distributed. However, if there are no constraints on the values of $\beta_i$ and $\tau_i$, the model becomes difficult to fit, because for example there would be no difference in data generated by $\beta_i=3, \tau_i=3$ and $\beta_i=3000, \tau_i=3000$ for a particular character $i$, regardless of the value $\lambda_i$. Because this makes inference problematic, we assume that the $\beta$ and $\tau$ distributions are truncated in the interval $[0,7]$. The overall model is:

\subsection{Model. }\label{model}
$$X_{it} \sim
\begin{cases}
\mathrm{Pois}(\lambda_i) &\qquad\text{if } |t-\beta_i| < \tau_i\\
0 &\qquad \text{otherwise.}
\end{cases}
$$

for $1 \le i \le N$, and $t \in \{1, 2, 3, 4, 5, 6, 7\}$, with

\begin{align*}
\log(\lambda_i) &\sim N(\mu_\lambda, \sigma_\lambda^2)\\
\tau_i &\sim N(\mu_\tau, \sigma_\tau^2) \text{ truncated to } [0,7]\\
\beta_i &\sim N(\mu_\beta, \sigma_\beta^2) \text{ truncated to } [0,7]
\end{align*}

where $\sigma_\lambda, \sigma_\tau, \sigma_\beta > 0$ and $\mu_\lambda, \mu_\tau, \mu_\beta \in \mathbb{R}$. For fixed $i$, the $X_{it}$ are assumed to be conditionally independent given $\lambda_i, \tau_i$ and $\beta_i$. For fixed $t$ and $i \neq j$, the $X_{it}$ and $X_{jt}$ are assumed to be conditionally independent given the values of $\lambda_i, \tau_i, \beta_i$ and $\lambda_j, \tau_j$ and $\beta_j$.

\subsection{}To be explicit, let $(x_{it})_{1 \le i \le N}^{1 \le t \le d}$ be the data. For $1 \le i \le N$ define $L_i =$ $L_i((x_{it})_{t=1}^d, \lambda_i, \tau_i, \beta_i)$ by
$$L_i = \prod_{t: x_{it}\neq 0}\frac{e^{-\lambda_i}\lambda_i^{x_{it}}}{x_{it}!}\prod_{t: x_{it}=0}
(e^{-\lambda_i}\delta_{|t-\beta_i|<\tau_i} + \delta_{|t-\beta_i|\ge \tau_i}).$$
Then the likelihood is proportional to
\begin{equation}\label{likelihood}
\int \prod_{i=1}^N L_i \frac{1}{\sigma_\lambda}e^{-\frac{(\log(\lambda_i)-\mu_\lambda)^2}{2\sigma_\lambda^2}}
\frac{1}{\sigma_\tau}e^{-\frac{(\tau_i-\mu_\tau)^2}{2\sigma_\tau^2}}
\frac{1}{\sigma_\beta}e^{-\frac{(\beta_i-\mu_\beta)^2}{2\sigma_\beta^2}}
\delta_{0\le \tau_i \le 7}\delta_{0 \le \beta_i \le 7}
\end{equation}
where the integral is over the $3N$ dimensions $\lambda_i,\tau_i,\beta_i \ge 0$ and the symbol $\delta_p$ stands for $1$ if $p$ is true and $0$ if $p$ is false.

\subsection{}A model like \ref{model} is often called a \emph{hierarchical} model. The $\mu_\lambda, \sigma_\lambda, \mu_\tau, \sigma_\tau, \mu_\beta$ and $\sigma_\beta$ are called \emph{hyperparameters} to distinguish them from the individual $\lambda_i$, $\tau_i$ and $\beta_i$.

\subsection{Method of inference. }The model is fitted using Bayesian inference with non-informative $N(0,1000^2)$ priors on the location parameters $\mu_\lambda$, $\mu_\tau$ and $\mu_\beta$ and inverse gamma $(0.001, 0.001)$ priors on the scale parameters $\sigma_\lambda, \sigma_\tau$ and $\sigma_\beta$. Because intractable-looking integrals appear in the likelihood (\ref{likelihood}), the model is fitted using Gibbs sampling. For the $\lambda_i$ and $\beta_i$, samples are drawn from the marginal distribution using a histogram approximation. This is slow but easier to code than alternatives.

\subsection{}\label{algorithm}At each iteration of the algorithm, a value of $(\mu_\lambda, \sigma_\lambda, \mu_\tau, \sigma_\tau, \mu_\beta, \sigma_\beta)$ is sampled using the theory of the normal distribution and then, for each character $i$, the values of $\lambda_i$, $\tau_i$ and $\beta_i$ are sampled in that order. Then predictions for $X_{i,d+1}$ and $X_{i, d+2}$ are sampled using the definition of Model \ref{model}. After all iterations are complete, a burn-in is discarded and the output is thinned to make the resulting samples as uncorrelated as possible. This is useful for some purposes, such as drawing Figure \ref{prob0}.

\subsection{}The output of the algorithm is a collection of samples of $(\mu_\lambda, \sigma_\lambda, \mu_\tau, \sigma_\tau, \mu_\beta, \sigma_\beta)$ and predictions $(x_{i, d+1})_{i=1}^n$ and $(x_{i, d+2})_{i=1}^n$ where $n = (N-\text{ burn-in})/\text{thin}$. The algorithm is written in $\mathsf{R}$ \cite{R} and uses the \texttt{truncnorm} package \cite{truncnorm}.

\section{Results}
\subsection{Data smoothing. }\label{datasmoothing} Model \ref{model} does not fit the training data very well since there are so many zeroes in column AFFC in Table \ref{M}. It is known \cite[afterword]{AFFC}, \cite[foreword]{ADWD} that \cite{AFFC} and \cite{ADWD} were originally planned to be a single book, but it was later split into two volumes, each of which concentrates on a different subset of the characters.

\subsection{}\label{preprocess}This problem can be approached either by ignoring it, by modelling, or by pre-processing the data. The model already has a lot of parameters and making it more complex is unlikely to be a good idea. Ignoring the problem and fitting the model to Table \ref{M} is not too bad, but it was decided to pre-process the data by replacing $M$ by $M'$ where
\begin{align*}
M'_{i4} &= c_4 \frac{M_{i4}+M_{i5}}{c_4+c_5}\\
M'_{i5} &= c_5 \frac{M_{i4}+M_{i5}}{c_4+c_5}
\end{align*}
where $c_4$ and $c_5$ are the number of chapters in books $4$ and $5$ respectively. This preserves the total number of chapters in each book, which may be of interest (see Section \ref{newcharacters}.)

\subsection{} Another possible approach would be to treat books $4$ and $5$ as one giant book when fitting the model. The main disadvantage of this approach is that it decreases the amount of data available even further, although the resulting $24 \times 4$ matrix would probably provide a better fit than $M'$ to the chosen model.

\subsection{Posterior predictive distributions. }The Gibbs sampler of Section \ref{algorithm} was run for $101000$ iterations with a burn-in of $1000$ and thinned by taking every $100^{\mathrm{th}}$ sample, resulting in posterior samples of size $n=1000$. The algorithm was applied to the smoothed data $M'$ of section \ref{datasmoothing}. It was run several times with random starting points to check that the results are stable. Only one run is recorded here.

\subsection{}
\begin{figure}[ht]
\centering
\includegraphics[scale=0.29]{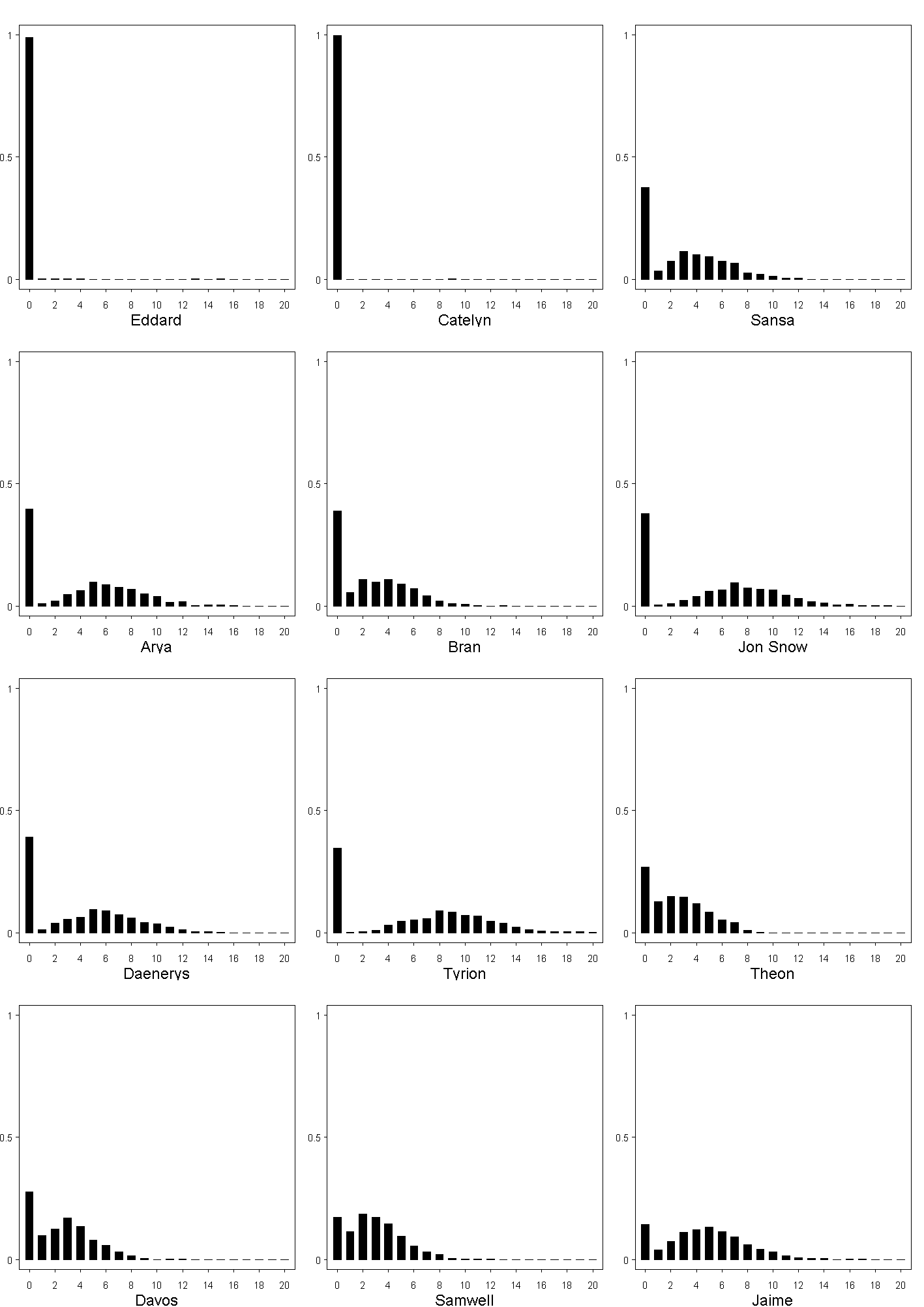}
\caption{Posterior predictive distributions for the number of POV chapters for twelve characters in \emph{The Winds of Winter}.}\label{posteriors_1}
\end{figure}

\begin{figure}[ht]
\centering
\includegraphics[scale=0.29]{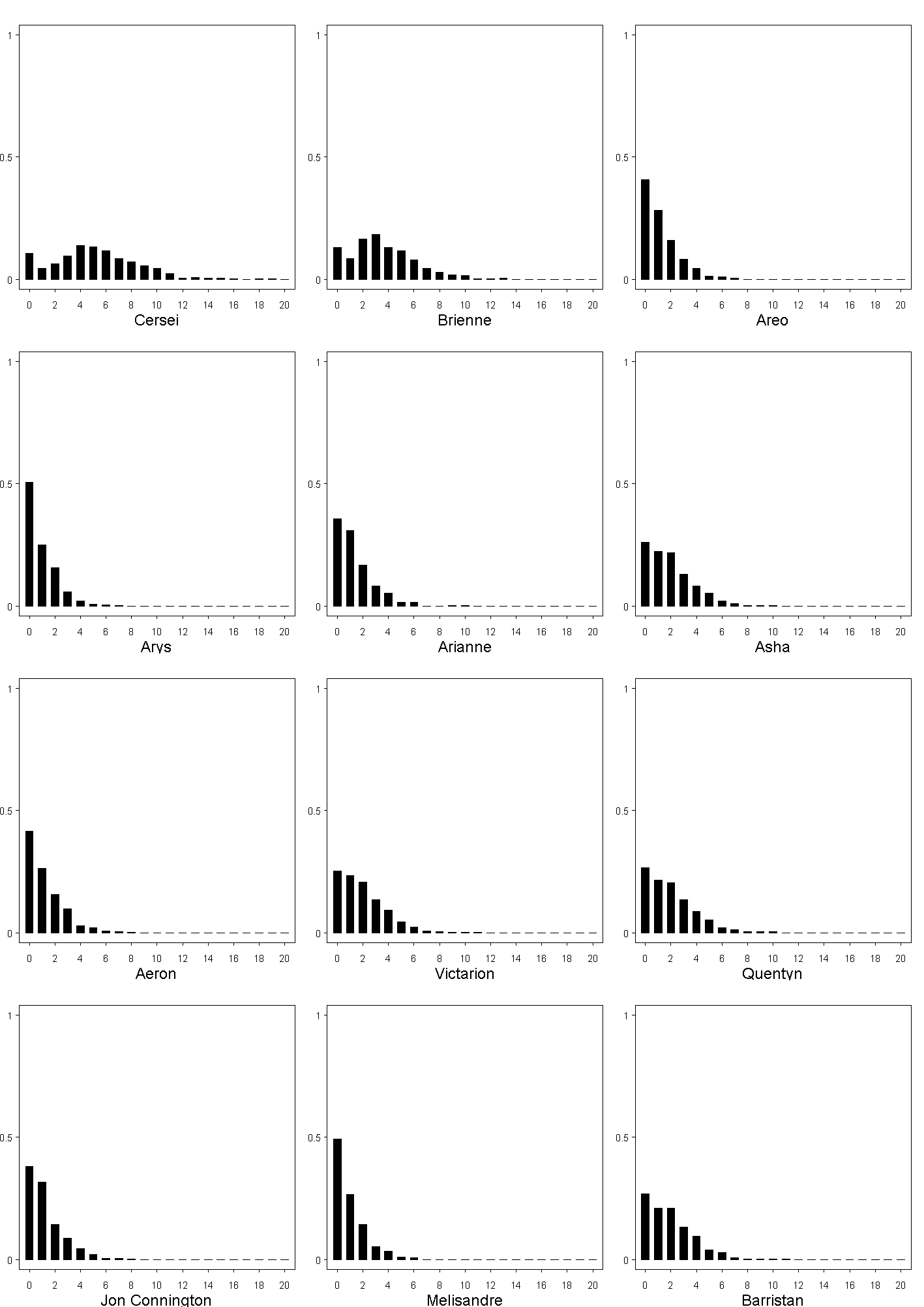}
\caption{Posterior predictive distributions for the number of POV chapters for twelve characters in \emph{The Winds of Winter}. Within the subsets $\{$Aeron, Areo, Jon Connington, Arianne$\}$, $\{$Quentyn, Victarion, Asha, Barristan$\}$ and $\{$Arys, Melisandre$\}$ the distributions are identical; apparent differences are due to sampling variation. }\label{posteriors_2}
\end{figure}

Table \ref{posteriors1} gives the posterior predictive distribution of POV chapters for each POV character in book $6$. Table \ref{posteriors2} gives a similar distribution for book $7$. (The results for book $7$ are of less interest, as new predictions for book $7$ should be made following the appearance of book $6$.) Graphs of the posterior distributions for book $6$ are given in Figures \ref{posteriors_1} and \ref{posteriors_2}. Many of the distributions are bimodal and are not well-summarised by a single credible interval. The distribution for Tyrion has the highest variance, followed by Jon Snow (see Figure \ref{posteriors_1}.)

\afterpage{
\begin{landscape}
\mbox{}\vfill
\begin{table}[!ht] 
\begin{threeparttable}
\begin{tabular}{l|ccccccccccccccccccccccc}
& 0& 1 &2& 3&4& 5& 6& 7& 8& 9& 10& 11& 12& 13& 14& 15& 16& 17& 18& 19& 20 &21&22\\
\hline
Eddard& 991& 2& 1& 2& 2& 0& 0& 0& 0& 0& 0& 0& 0& 1& 0& 1& 0& 0& 0& 0& 0& 0& 0\\
Catelyn& 999& 0& 0& 0& 0& 0& 0& 0& 0& 1& 0& 0& 0& 0& 0& 0& 0& 0& 0& 0& 0& 0& 0\\
Sansa& 377& 34& 74& 114& 100& 93& 74& 66& 26& 20& 13& 5& 4& 0& 0& 0& 0& 0& 0& 0& 0& 0& 0\\
Arya& 397& 10& 21& 46& 62& 99& 87& 76& 68& 50& 40& 16& 17& 1& 5& 4& 1& 0& 0& 0& 0& 0& 0\\
Bran& 389& 55& 108& 98& 108& 91& 71& 42& 20& 10& 6& 1& 0& 1& 0& 0& 0& 0& 0& 0& 0& 0& 0\\
Jon Snow& 378& 4& 9& 22& 39& 60& 66& 95& 75& 67& 65& 44& 30& 18& 13& 5& 6& 2& 1& 1& 0& 0& 0\\
Daenerys& 393& 12& 40& 55& 63& 94& 90& 74& 59& 41& 37& 23& 11& 3& 3& 2& 0& 0& 0& 0& 0& 0& 0\\
Tyrion& 347& 2& 3& 9& 30& 48& 53& 57& 91& 84& 70& 67& 46& 38& 22& 11& 7& 4& 4& 4& 1& 1& 1\\
Theon& 269& 128& 149& 146& 119& 84& 51& 42& 10& 2& 0& 0& 0& 0& 0& 0& 0& 0& 0& 0& 0& 0& 0\\
Davos& 278& 97& 125& 171& 136& 80& 57& 32& 16& 5& 0& 2& 1& 0& 0& 0& 0& 0& 0& 0& 0& 0& 0\\
Samwell& 172& 114& 186& 172& 146& 94& 54& 32& 20& 5& 1& 2& 2& 0& 0& 0& 0& 0& 0& 0& 0& 0& 0\\
Jaime& 144& 40& 74& 111& 121& 133& 113& 93& 61& 42& 32& 15& 8& 5& 5& 0& 2& 1& 0& 0& 0& 0& 0\\
Cersei& 105& 45& 64& 96& 139& 132& 117& 84& 72& 55& 45& 23& 5& 7& 4& 3& 1& 0& 2& 1& 0& 0& 0\\
Brienne& 130& 85& 165& 184& 131& 116& 79& 43& 28& 18& 14& 2& 1& 4& 0& 0& 0& 0& 0& 0& 0& 0& 0\\
Areo& 408& 281& 159& 82& 45& 13& 9& 3& 0& 0& 0& 0& 0& 0& 0& 0& 0& 0& 0& 0& 0& 0& 0\\
Arys& 508& 249& 156& 57& 19& 6& 4& 1& 0& 0& 0& 0& 0& 0& 0& 0& 0& 0& 0& 0& 0& 0& 0\\
Arianne& 358& 310& 166& 83& 51& 15& 14& 0& 0& 2& 1& 0& 0& 0& 0& 0& 0& 0& 0& 0& 0& 0& 0\\
Asha& 262& 223& 219& 129& 81& 51& 20& 10& 2& 2& 1& 0& 0& 0& 0& 0& 0& 0& 0& 0& 0& 0& 0\\
Aeron& 417& 263& 157& 99& 29& 21& 8& 4& 2& 0& 0& 0& 0& 0& 0& 0& 0& 0& 0& 0& 0& 0& 0\\
Victarion& 252& 234& 206& 136& 92& 43& 22& 6& 4& 2& 2& 1& 0& 0& 0& 0& 0& 0& 0& 0& 0& 0& 0\\
Quentyn& 265& 215& 205& 136& 86& 52& 19& 12& 4& 3& 3& 0& 0& 0& 0& 0& 0& 0& 0& 0& 0& 0& 0\\
Jon C& 380& 316& 144& 88& 45& 19& 3& 4& 1& 0& 0& 0& 0& 0& 0& 0& 0& 0& 0& 0& 0& 0& 0\\
Melisandre& 492& 265& 143& 52& 33& 9& 6& 0& 0& 0& 0& 0& 0& 0& 0& 0& 0& 0& 0& 0& 0& 0& 0\\
Barristan& 270& 211& 209& 133& 94& 40& 29& 8& 2& 1& 2& 1& 0& 0& 0& 0& 0& 0& 0& 0& 0& 0& 0
\end{tabular}
\vskip 2ex
\caption{1000 posterior samples of the number of POV chapters for each character in \emph{The Winds of Winter}.}
\label{posteriors1}
\end{threeparttable}
\end{table}
\vfill
\end{landscape}
}

\afterpage{
\begin{landscape}
\mbox{}\vfill
\begin{table}[!ht] 
\begin{threeparttable}
\begin{tabular}{l|ccccccccccccccccccccccc}
& 0& 1 &2& 3&4& 5& 6& 7& 8& 9& 10& 11& 12& 13& 14& 15& 16& 17& 18& 19& 20 &21&22\\
\hline
Eddard& 996& 0& 0& 1& 2& 0& 1& 0& 0& 0& 0& 0& 0& 0& 0& 0& 0& 0& 0& 0& 0& 0& 0\\
Catelyn& 999& 0& 0& 0& 0& 0& 0& 1& 0& 0& 0& 0& 0& 0& 0& 0& 0& 0& 0& 0& 0& 0& 0\\
Sansa& 654& 16& 41& 61& 65& 46& 41& 35& 25& 8& 5& 2& 1& 0& 0& 0& 0& 0& 0& 0& 0& 0& 0\\
Arya& 700& 4& 8& 28& 38& 39& 47& 46& 26& 23& 19& 8& 7& 5& 1& 0& 1& 0& 0& 0& 0& 0& 0\\
Bran& 682& 26& 46& 49& 58& 65& 32& 19& 13& 5& 3& 2& 0& 0& 0& 0& 0& 0& 0& 0& 0& 0& 0\\
Jon Snow& 672& 1& 4& 16& 22& 29& 43& 38& 33& 35& 48& 17& 18& 10& 8& 4& 2& 0& 0& 0& 0& 0& 0\\
Daenerys& 678& 7& 15& 39& 49& 47& 44& 33& 26& 22& 18& 7& 7& 2& 1& 5& 0& 0& 0& 0& 0& 0& 0\\
Tyrion& 653& 0& 4& 9& 15& 21& 30& 44& 40& 42& 33& 31& 30& 20& 11& 6& 3& 2& 3& 2& 0& 1& 0\\
Theon& 542& 61& 97& 115& 80& 45& 24& 19& 5& 9& 3& 0& 0& 0& 0& 0& 0& 0& 0& 0& 0& 0& 0\\
Davos& 552& 56& 80& 91& 93& 57& 35& 21& 7& 7& 1& 0& 0& 0& 0& 0& 0& 0& 0& 0& 0& 0& 0\\
Samwell& 410& 93& 123& 133& 96& 61& 40& 20& 18& 5& 1& 0& 0& 0& 0& 0& 0& 0& 0& 0& 0& 0& 0\\
Jaime& 417& 25& 53& 72& 94& 91& 74& 66& 47& 27& 14& 10& 3& 3& 3& 0& 1& 0& 0& 0& 0& 0& 0\\
Cersei& 284& 29& 55& 76& 104& 88& 101& 84& 70& 36& 24& 27& 10& 7& 1& 3& 0& 1& 0& 0& 0& 0& 0\\
Brienne& 278& 77& 133& 140& 134& 94& 58& 43& 24& 11& 4& 3& 0& 1& 0& 0& 0& 0& 0& 0& 0& 0& 0\\
Areo& 537& 217& 124& 59& 34& 19& 7& 0& 3& 0& 0& 0& 0& 0& 0& 0& 0& 0& 0& 0& 0& 0& 0\\
Arys& 601& 211& 105& 55& 20& 6& 2& 0& 0& 0& 0& 0& 0& 0& 0& 0& 0& 0& 0& 0& 0& 0& 0\\
Arianne& 537& 222& 126& 59& 26& 14& 12& 3& 1& 0& 0& 0& 0& 0& 0& 0& 0& 0& 0& 0& 0& 0& 0\\
Asha& 432& 183& 162& 120& 56& 23& 14& 6& 3& 1& 0& 0& 0& 0& 0& 0& 0& 0& 0& 0& 0& 0& 0\\
Aeron& 575& 185& 127& 62& 24& 19& 7& 1& 0& 0& 0& 0& 0& 0& 0& 0& 0& 0& 0& 0& 0& 0& 0\\
Victarion& 408& 173& 178& 103& 60& 33& 26& 11& 4& 4& 0& 0& 0& 0& 0& 0& 0& 0& 0& 0& 0& 0& 0\\
Quentyn& 423& 172& 172& 110& 59& 44& 12& 4& 2& 2& 0& 0& 0& 0& 0& 0& 0& 0& 0& 0& 0& 0& 0\\
Jon C& 575& 182& 130& 61& 31& 12& 6& 2& 0& 1& 0& 0& 0& 0& 0& 0& 0& 0& 0& 0& 0& 0& 0\\
Melisandre& 598& 211& 117& 45& 16& 6& 6& 1& 0& 0& 0& 0& 0& 0& 0& 0& 0& 0& 0& 0& 0& 0& 0\\
Barristan& 415& 184& 162& 110& 77& 27& 16& 7& 1& 1& 0& 0& 0& 0& 0& 0& 0& 0& 0& 0& 0& 0& 0\\
\end{tabular}
\vskip 2ex
\caption{1000 posterior samples of the number of POV chapters for each character in the book following \emph{The Winds of Winter}.}
\label{posteriors2}
\end{threeparttable}
\end{table}
\vfill
\end{landscape}
}

\subsection{Probabilities for zero POV chapters. }\label{zeroprobabilities}
\begin{figure}[!htbp]
\centering
\includegraphics[scale=0.7]{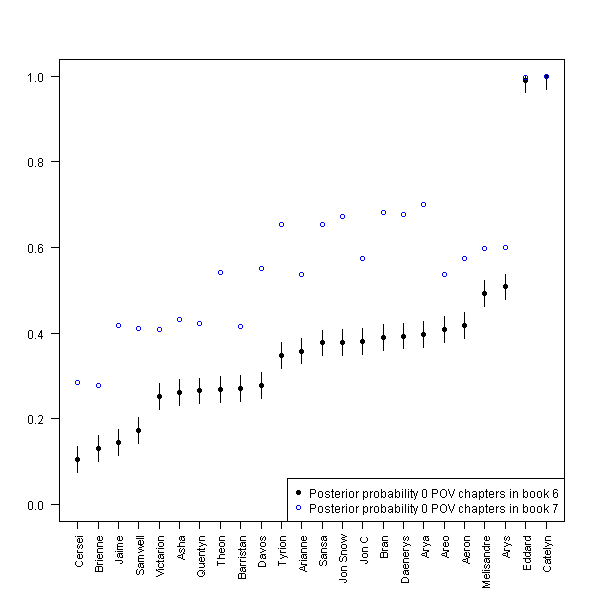}
\caption{Posterior probabilities for each of the existing POV characters to have zero POV chapters in the next book. These probabilities correspond to the entries in the first column of Table \ref{posteriors1}. The characters are ordered on the $x$--axis by the value of the posterior probability for book $6$ (the black dots.) The blue circles are the posterior probabilities of having zero POV chapters in book $7$. The error bars extend to $\pm 3\%$ and are intended to indicate when two posterior probabilities are roughly equal. The Figure is discussed in Sections \ref{zeroprobabilities} to \ref{isjonsnowdead}.}\label{prob0}
\end{figure}

One of the most compelling aspects of the \emph{Song of Ice and Fire} series is that major characters are frequently and unexpectedly killed off. The probability of a character having zero POV chapters (which is the first column of Table \ref{posteriors1} divided by $1000$) is therefore of interest. These values are plotted in Figure \ref{prob0}. Treating the posterior samples as independent, the error bars in Figure \ref{prob0} indicate approximate $95\%$ confidence intervals of $\pm 3\%$. Note that although a character who has been killed off will have zero POV chapters, the converse is not necessarily true. The probabilities in Figure \ref{prob0} are not based on events in the books, but solely reflect what the model can glean from Table \ref{M}.

\subsection{}The characters in Figure \ref{prob0} are arranged on the $x$--axis in order of their probability of having zero POV chapters in book $6$. Eddard and Catelyn were already killed off in \cite{AGOT} and (arguably) \cite{ASOS} respectively. Arys, who has the third highest posterior probability of having zero POV chapters, was killed off in \cite{AFFC}, but it is misleading to be impressed by this. The reason why Arys has a high posterior probability of having zero POV chapters is that he only ever appeared in one POV chapter, so $\lambda_{\text{Arys}}$ is small and so, even if $\tau_{\text{Arys}}$ is large, there is a high probability that $X_{\text{Arys}, 6} = 0$. In fact, in the smoothed data, the rows for Melisandre and Arys are exactly the same. The difference between the posterior distributions for these two characters is due to sampling variation (and varies from one model run to another.)

\subsection{}The characters between Tyrion and Aeron in Figure \ref{prob0} all have roughly the same posterior probability of having zero POV chapters. They are mostly characters who have featured prominently in all the books since the beginning, together with the newer characters Aeron, Areo, Arianne and Jon Connington whose posterior predictive distributions are identical and who have lower probabilities of non-appearance in book $7$.

\subsection{}The next group of characters are those who have had relatively few POV chapters, including Quentyn who, despite having been killed off in \cite{ADWD}, is assigned a low posterior probability of $0.265$ of having zero POV chapters in book $6$ since he has the same posterior predictive distribution as Asha, Barristan and Victarion.

\subsection{}Finally, there are the characters Cersei, Brienne, Jaime and Samwell who have only recently become POV characters and have had a large number of POV chapters in the books in which they have appeared.

\subsection{Is Jon Snow dead? }\label{isjonsnowdead}The model suggests that the probability of Jon Snow \emph{not} being dead is at least $60\%$ since this is less than the posterior probability of his having at least one POV chapter in book $6$. Given the events of \cite{ADWD}, many readers would assess his probability of not being dead as being much lower than $60\%$, but we must again point out that the model is unaware of the events in the books. The model can only say that, based on the number of POV chapters observed so far, he has about as much chance of survival as the other major characters.

\section{Testing and Validation}

\subsection{Testing the method of inference. }It is desirable to check that the model has been coded correctly. A way to check this is to generate a data set according to the model and then see whether the chosen method of inference can recover the parameters which were used to generate the data set.

\subsection{}If the model had been fitted by frequentist methods, it would be possible to generate a large number of data sets, fit the model to each one, calculate confidence intervals for the hyperparameters, and check that the confidence intervals have the correct coverage. Since the model has been fitted by Bayesian methods, it can only be used to produce credible intervals. However, as flat priors were used, the posterior distributions for the hyperparameters should be close to their likelihoods and so Bayesian credible intervals should roughly coincide with frequentist confidence intervals when the posterior distributions of the hyperparameters are symmetric and unimodal, which they are.

\subsection{}\label{inferencetest}
\begin{figure}[!htbp]\label{inference_test}
\centering
\includegraphics[scale=0.7]{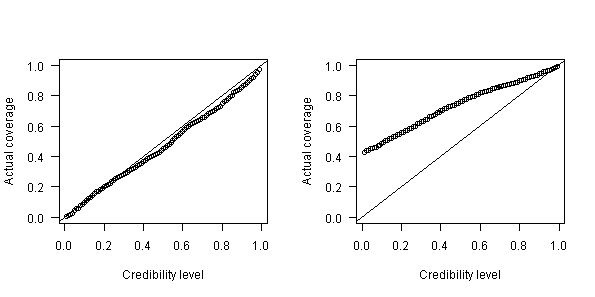}
\caption{Results of testing the method of inference on $100$ model fits as described in Section \ref{inferencetest} and Section \ref{inferencetest_pred}. On the left, a comparison of the nominal and actual coverage of credible intervals for the hyperparameters. On the right, a comparison of the nominal and actual coverage of credible intervals for the predicted number of POV chapters.}
\end{figure}

To test the method of inference, the model was fitted to $100$ data sets. Each data set was generated from hyperparameters which were a perturbation of $(\mu_\lambda, \sigma_\lambda, \mu_\tau, \sigma_\tau, \mu_\beta, \sigma_\beta) = (1.3, 0.75, 2, 1, 4, 1.5)$. These values were chosen because they were approximately the posterior medians for the hyperparameters obtained from one of the fits of Model \ref{model} to the smoothed data $M'$. The location parameters $\mu_\lambda, \mu_\tau$ and $\mu_\beta$ were perturbed by adding $N(0, 0.1^2)$ noise and the scale parameters $\sigma_\lambda, \sigma_\tau$ and $\sigma_\beta$ were perturbed by multiplying by $\exp(\Delta)$ where $\Delta \sim N(0, 0.01^2)$. For each of the data sets, $\alpha$--credible intervals were calculated by taking the middle $100\alpha\%$ of the posterior distributions for each of the $6$ hyperparameters, yielding $600$ credible intervals per $\alpha$. The results, plotted in Figure \ref{inference_test} (left panel) indicate that the credible intervals have roughly the correct coverage.

\subsection{}Note that there are choices of the hyperparameters which the chosen method of inference will not be able to recover. For example, data generated with $\mu_\beta=1000$ will be practically indistinguishable from data generated with $\mu_\beta=100$, so there is no hope of inferring the hyperparameters in this case. This is no great drawback as it should not affect the model's predictions, which are the topic of interest.

\subsection{}\label{inferencetest_pred}The procedure of Section \ref{inferencetest} was carried out for the one-step-ahead predictions for book $6$. The result, shown in Figure \ref{inference_test} (right panel) shows that the credible intervals have greater coverage than they should. This is because the number of POV chapters can only take integer values and so the closed interval $[q_{\alpha/2}, q_{1-\alpha/2}]$ obtained by taking the $\alpha/2$ and $1-\alpha/2$ quantiles of the posterior distribution will in general cover more than $100\alpha\%$ of the posterior samples.

\subsection{}Note once again that the purpose of these checks and experiments is to make sure that the model has been correctly coded. We now discuss how to evaluate its predictions.

\subsection{Validation. }Every predictive model should be applied to unseen test data to see how accurate its predictions really are. It will not be possible to test Model \ref{model} before the publication of \cite{TWOW} but an attempt at validation can be made by fitting the model to earlier books and seeing what it would have predicted for the next book.

\subsection{}\label{validation_12}
\begin{figure}[!htbp]\label{validation}
\centering
\includegraphics[scale=0.5]{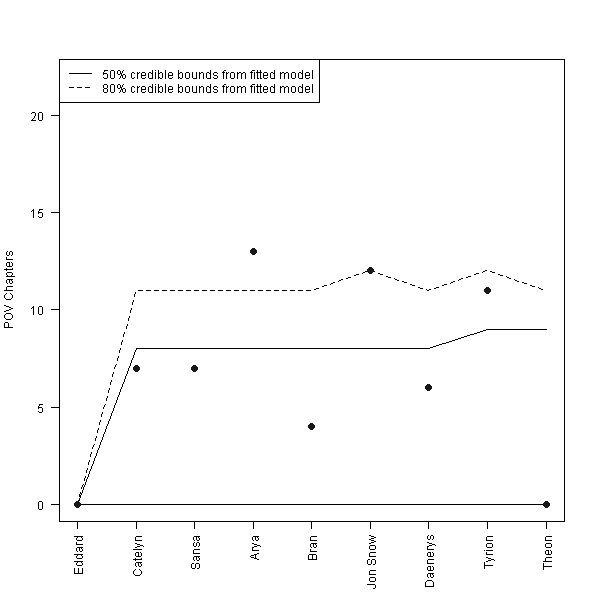}
\caption{Actual values of $M_{i3}, 1 \le i \le 9$ (plotted as dots) from Table \ref{M} compared with central $50\%$ (solid lines) and $80\%$ (broken lines) credible intervals obtained from fitting the model to $(M_{ij})_{1\le i \le 9, 1\le j \le 2}$ as described in Section \ref{validation_12}. The characters have been sorted in increasing order of the posterior median.}
\end{figure}

The model was tested by fitting it to  books $1$ and $2$ in the series. Only $9$ POV characters appear in these books, so the data consist of the upper-left $9\times 2$ submatrix of Table \ref{M}. Figure \ref{validation} shows the result of fitting the model to this matrix and comparing with the true values from the third column of Table \ref{M}. The intervals displayed are central $50\%$ (solid lines) and $80\%$ (dotted lines) credible intervals. The coverage is satisfactory but the intervals are much too wide to be of interest.

\subsection{}Fitting the model to the $12 \times 3$ upper-left submatrix of Table \ref{M} consisting of POV chapters from the first three books gives more interesting output, but it is not clear how to evaluate the results because of the splitting of books $4$ and $5$ discussed above in Section \ref{datasmoothing}.

\subsection{}We can also compare the model's predictions with preview chapters from \cite{TWOW} which are said to have been released featuring the points of view of Arya, Arianne, Victarion and Barristan. Given that there is at least one Arya chapter, Table \ref{posteriors1} indicates that there will probably be at least $5$ Arya POV chapters and perhaps more.

\section{Issues with the model}

\subsection{}Given that we are interested in whether the model works for its intended purpose rather than in advertising it, we should not shy away from identifying and criticising its flaws.

\subsection{Zero histories. }The model can generate data containing a row of zeroes, but there are no zero rows in the data to which the model is fitted, because by definition this would correspond to a character who has never been a POV character in the books. This is a source of bias in the model but it is not obvious how it can be avoided. The effect of the bias can be tested by repeating the simulations of Section \ref{inferencetest}, but deleting zero rows before fitting the model. For $0.5 \le \alpha \le 0.95$, the coverage of a $100\alpha\%$ credible interval for a hyperparameter tends to be roughly $\alpha-0.1$.

\subsection{Poisson assumption. }There is little to support the choice of the Poisson distribution in Model \ref{model} other than that it has the smallest possible number of parameters. It is more common to use the negative binomial distribution for count data, but this would introduce extra complexity into the model, which is undesirable.

\subsection{Not enough data. }With only $24$ values of $\lambda_i$, $\tau_i$ and $\beta_i$ available for finding the corresponding hyperparameters, it may not be possible to fit a (truncated) normal distribution in a meaningful way. Consideration of the posterior samples of the $\lambda_i$, $\tau_i$ and $\beta_i$ suggest that they more-or-less follow the pattern which is evident in the data and that the shrinkage of these parameters towards a common mean, which is one of the benefits of using a hierarchical model, cannot really be attained with so little data. For example, when the model is fitted to a data set containing a row in which the most recent entry is $0$, the vast majority of posterior samples for the next entry in that row are always $0$. This is one reason for smoothing the data in Section \ref{datasmoothing} before fitting the model, in preference to fitting the model directly to Table \ref{M}. If the number of POV characters was much larger, this might not be such a big problem.

\subsection{Lack of independence. }Given $\lambda_i, \tau_i, \beta_i, \lambda_j, \tau_j, \beta_j$, the model treats $X_{it}$ and $X_{jt}$ as independent for $i \neq j$. This is not a realistic assumption because if one character has more POV chapters, then the other characters will necessarily have fewer. Again, addressing this would seem to over-complicate the model.

\subsection{New characters. }\label{newcharacters}The model ignores the introduction of new POV characters, although every book in the series has featured some new POV characters. We can, however, use the output from the fitted model to make guesses about new characters. The posterior distribution of the number of chapters in book $6$ told from the points of view of existing POV characters is unimodal with a mean of $59.3$ chapters, but typical books in the series so far have had about $70$ chapters. So we could estimate that there will be about $11$ chapters in \cite{TWOW} told from the point of view of new POV characters. In the previous books, according to Table \ref{M}, the number of chapters told from the point of view of new POV characters has been $9, 14, 27$ and $11$, so $11$ does not seem like an unreasonable guess.

\subsection{}We could continue to make predictions in the hope of getting one right, but there is no merit in this. We hope that it will be possible to review the model's performance following the publication of \cite{TWOW}.


\end{document}